\documentstyle[11pt]{article}
\newcommand{\nc}{\newcommand}
\nc{\zm}[1]{\stackrel{o} {#1} }
\nc{\nm}[1]{\stackrel{n} {#1} }
\nc{\dslash}{\not{\! \partial}}
\nc{\Dslash}{\not{\!\! D}}
\nc{\be}{\begin{equation}}
\nc{\ee}{\end{equation}}
\nc{\noi}{\noindent}
\begin{document} \thispagestyle{empty}
\setcounter{page}{0}
\noindent\hfill  \\
\noindent\hfill  OHSTPY-HEP-TH-94-001  \\
\noindent\hfill  MPIH-V9-94\\
\noindent\hfill 27 May, 1994 \\
\begin{center}\begin{Large}\begin{bf}
Dynamical Zero Modes and Pure Glue ${\bf{\rm{QCD}}_{1+1}}$ \\
in Light-Cone Field Theory \\
\end{bf}\end{Large}\end{center}
\vspace{0.5cm}\begin{center}
Alex C. Kalloniatis and Hans-Christian Pauli\\[10pt]
      \end{center}
     \begin{center}
     \begin{it}
Max-Planck-Institut fur Kernphysik\\
Postfach 10 39 80\\
D-69029 Heidelberg 1\\
       \end{it}
        \end{center}
        \vspace{0.5cm}
      \begin{center}
Stephen Pinsky \\[10pt]
     \end{center}
      \begin{center}
      \begin{it}
Department of Physics\\
The Ohio State University\\
174 West 18th Avenue\\
Columbus, Ohio  43210\\
       \end{it}
         \end{center}
\vspace{0.5cm}
\begin{abstract} \noindent
We consider light-cone quantized ${\rm{QCD}}_{1+1}$ on a `cylinder'
with periodic boundary conditions on the gluon fields. This
is the framework of discretized light-cone quantization.
We review the argument that
the light-cone gauge $A^+=0$ is not attainable. The zero
mode is a dynamical and gauge invariant field.
The attainable gauge has a Gribov ambiguity.
We exactly solve the problem of pure glue theory coupled to
some zero mode external sources. We verify the identity of the
front and the more familiar instant form
approaches.
We obtain a discrete spectrum of vacuum states and their wavefunctions.
\end{abstract}
\newpage
\section{Introduction}
Recently the Hamiltonian approach to field theory has been tackled with
renewed interest. The hope is that Dirac's `front form' Hamiltonian scheme
\cite{Dir49}
is useful for confronting quantum chromodynamics (QCD). Often in the literature
this is called `light-cone', `null-plane' or `light-front' quantization. In the
sequel we shall persist with the original Dirac nomenclature.
This formulation uses
$\displaystyle{x^+ = {1\over{\sqrt{2}}}(c t + z)}$, called the light-cone time,
as the `time' evolution parameter rather than the conventional $x^0 = c t$.
For an extensive bibliography the reader is referred to Refs. \cite{PW}.
One reason for the modern phase of this approach is the apparent simplicity of
the vacuum
in front form theory. In the more familiar `instant form' quantization the
QCD vacuum contains an infinite number of soft particles.
But then in front form field theory the question arises: where can
long range phenomena of spontaneous symmetry breaking and
perhaps even confinement
appear in the apparent absence of any
`infrared' vacuum structure?

The specific approach of {\it Discretized} Light-Cone Quantization (DLCQ) is
one setting in which one can anwer these questions and hopefully pursue the
program to a solution. Here the theory is defined in a finite
`spatial volume' with periodic or antiperiodic
boundary conditions imposed on bosonic or fermionic fields, respectively.
There are two appealing reasons for such a formulation.
One obtains an infrared
regulated theory, and the discretization of momenta facilitates putting
the many-body problem onto the computer. The price one has to pay, shown
actually
some time ago \cite{Mas76}, is that Fourier zero modes of the fields
are often not independent dynamical quanta. Rather, by a
constraint equation, they are dependent on them.
Recent work on such a constrained zero mode
in scalar $\phi^4_{1+1}$ has lead to the insight that
it gives rise to the phenomena of spontaneous symmetry breaking and field
condensates
\cite{zmconstraint},
aspects normally attributed to non-trivial vacuum structure.

Our concern in this paper, however, is with zero modes that are
true dynamical independent fields. One way they can arise is as follows.
Due to the boundary conditions in gauge theory one cannot fully implement the
traditional light-cone gauge $A^{+}=0$. The development of the
understanding of this problem in DLCQ can be traced in Refs. \cite{mccartor}.
The field $A^+$ turns out to have a zero mode which cannot be gauged away
\cite{Kall93}.
This mode is indeed dynamical, and is the object we study in this paper.
It has its analogue in instant form approaches to gauge theory.
For example, there exists a large body of work
on Abelian and non-Abelian
gauge theories in 1+1 dimensions quantized on
a cylinder geometry \cite{cyl}.
There indeed this dynamical zero mode plays an important role.

We too shall concern ourselves in the present work with non-Abelian gauge
theory
in 1+1 dimensions, revisiting the model introduced by 't Hooft \cite{tHo74}.
A DLCQ treatment of the theory,
giving meson and baryon spectra, and wavefunctions, was undertaken by
Hornbostel \cite{Horn90}. Apart from a modified approach by Lenz et al.
\cite{Lenz},
zero modes have been neglected in previous DLCQ studies of ${\rm{QCD}}_{1+1}$.
This we
rectify to some extent in the present paper.

The specific task we undertake here is to understand the zero mode {\it
subsector} of
the pure glue theory, namely where only zero mode external sources
excite only zero mode gluons. We shall see that this is not an approximation
but rather a consistent solution, a sub-regime within the complete
theory. A similar framing of the problem lies behind the work of L\"uscher
\cite{Lues83}
and van Baal \cite{vBa92} using the instant form Hamiltonian approach to pure
glue
gauge theory in 3+1 dimensions.
The beauty of this reduction in the 1+1 dimensional theory
is two-fold. First, it yields a theory which
is exactly soluble. This is useful given the dearth of soluble models in field
theory.
Secondly, the zero mode theory represents a paring down to the point
where the front and instant forms are manifestly {\it identical}, which is nice
to
know indeed.
We solve the theory in this specific dynamical regime and find a discrete
spectrum
of states whose wavefunctions can be completely determined. These states have
the
quantum numbers of the vacuum.
There is a summary and discussion of the results at the end of the paper.
The appendix explains notation.

\section{Gauge Fixing}
We consider an SU(2) non-Abelian gauge theory in 1+1 dimensions with classical
sources
coupled to the gluons. The Lagrangian density is
\be
{\cal L} = {1\over 2}\, {\rm Tr}\, (F_{\mu \nu} F^{\mu \nu}) + 2\, {\rm Tr}\,
(J_\mu A^\mu)
\ee
where
$F_{\mu \nu} = \partial _{\nu} A_{\nu} - \partial_{\nu} A_{\mu} -g[A_{\mu},
A_{\nu}]$.
With a finite interval in $x^-$ from $-L$ to $L$, we impose periodic boundary
conditions
on all gauge potentials $A_\mu$.

We now show that the light-cone gauge $A^+=0$ cannot be reached.
A gauge transformation $U$ bringing a gauge potential $B^\mu$, itself
in some arbitrary gauge configuration, to some other gauge configuration
$A^\mu$ is
\be
gA^{\mu} = \partial_{\mu} U U^{-1} +g U B^{\mu} U^{-1}
\;.
\label{gaugetrans}
\ee
Here $g$ is the coupling constant and
$U$ is an element of the Lie algebra
of SU(2). Clearly $U$ given by
\be
U = P \exp{[-g \int_{-L}^{x^-} {dy^- B^+(y^-)}] }
\label{naiveU}
\ee
will bring us to the gauge
$A{^+} = 0$.

We appear to have been successful in getting the light-cone gauge.
However, the element $U$ through which we wish to
achieve the gauge condition must
satisfy $Z_2$-periodic boundary conditions,
as in \cite{Het93}, namely
$U(x) = (\pm) U(x+2L)$.
Clearly Eq.(\ref{naiveU}) does not satisfy these boundary
conditions. So in fact the attempt has failed.

With the appendicial notation, a modification of Eq.(\ref{naiveU}) is
\be
U(x) = e^{gx^-\zm{B^+}} P e^{-g\int_{-L}^{x^-} dy^- B^+(y^-)}
\;.
\ee
Since $\zm{B^+}$ is the zero mode of $B^+$, this is an allowed gauge
transformation
but it does not completely bring us to the light-cone gauge.  We find
instead
\be
A^+  = \zm{B^+}
\;.
\ee
In other words, we cannot eliminate the zero mode of the gauge potential. The
reason is evident: it is {\it invariant} under periodic gauge
transformations. But of course we can always perform a rotation in color space.
In line with other authors \cite{Fran81}, we choose this so that $\zm{A_3^+}$
is
the only non-zero element, since in our representation only $\sigma^3$ is
diagonal.

In addition, we can impose the subsidiary gauge condition
\be
\zm{A^-_3} = 0
\;.
\label{tempcond}
\ee
The reason is that there still remains freedom to perform gauge transformations
that depend only on light-cone time $x^+$ and the color matrix $\sigma^3$.
The above condition Eq. (\ref{tempcond}) can be reached from the
arbitrary configuration $B^\mu$
by the Lie algebra element
\be
W = P \exp [- i g \int_{x_0^+}^{x^+} d\tilde{x}^+ \zm{B_3^-}(\tilde{x}^+)
{{\sigma^3} \over 2}  ]
\;,
\ee
where $x^+_0$ is some arbitrary but fixed light-cone time.
It, moreover, does not `undo' the previous gauge condition.

The above procedure would appear to have enabled complete fixing of the gauge.
This is still not so. Gauge transformations
\be
V = \exp\{i x^- ({{n\pi} \over {2L}}) {\sigma}^3\}
\label{GribU}
\ee
generate shifts, according to Eq.(\ref{gaugetrans}), in the zero mode component
\be
\zm{A^+_3} \rightarrow \zm{A^+_3} + {{n\pi}\over{gL}}
\;.
\ee
All of these possibilities, labelled by the integer $n$, of course still
satisfy
$\partial_- A^+=0$, but as one sees $n=0$ should not really be included.
One can verify that the transformations $V$ also preserve
the subsidiary condition, Eq.(\ref{tempcond}).
One notes that the transformation is
$x^-$-dependent and $Z_2$ periodic. It is
thus a simple example of a Gribov copy \cite{Grib78} in 1+1 dimensions.
We follow the conventional procedure by demanding
\be
\zm{A^+_3} \neq {n \pi \over gL}\;, \quad  n= \pm1, \pm2, \ldots
\;.
\ee
This eliminates singularity points at the Gribov `horizons' which
in turn correspond to a vanishing Faddeev-Popov determinant \cite{vBa92}.

\section{Equations of motion}
\noi
\underline{Equations for Pure Glue Theory.}
Ultimately, the argument that the vacuum in front form field theory is
trivial rests on
the {\it linearity} of the Euler-Lagrange equations of motion in the
light-cone time $x^+$. This itself stems from the expression for
the D'Alembertian in light-cone coordinates $\Box = \partial_+\partial_-$
in one space dimension. It is the very same fact that causes most zero modes
to be constrained when there are transverse
dimensions: the space derivative kills
the mode, thus eliminating the time derivative in the equation of motion.
However, a careful examination of the equations can sometimes reveal
double time derivatives $\partial_+^2$ due to the gauge structure. Thus
there can still be dynamical zero mode degrees of freedom even in DLCQ
which could, in principle, undermine the vacuum `triviality' argument.
This is what we now explore for SU(2).

The equations of motion for the theory are
\be
[D^{\mu}, F_{\mu \nu}] = \partial^{\mu} F_{\mu \nu} -g [A^{\mu},
 F_{\mu \nu}] = J_{\nu}
\;.
\ee
For our purposes it is convenient to break this equation up into
color components
$A{^{\mu} _a}$.  Color will always be the lower index.  Rather than the three
color fields $A{^{\mu} _1} , A{^{\mu} _2} ~{\rm and}~ A{^{\mu} _3}$ we will
use chiral notation with
$ A{^{\mu} _+} = A{^{\mu} _1}  + iA{^{\mu} _2}$ and
$A{^{\mu} _-} = A{^{\mu} _1} -iA{^{\mu} _2}$.
In terms of these components the equations of motion are
\begin{eqnarray}
\partial_\mu \partial^\mu A^\nu_3
 - \partial^\nu \partial_\mu A^\mu_3 & + & \nonumber \\
& & \makebox[-6cm]{}
 {ig \over 2} A^\mu_-
{\stackrel{\leftrightarrow} {\partial^\nu}}
 A_{\mu,+}
+ {ig \over2} (A^\nu_- \partial_\mu A^\mu_+ - A^\nu_+
  \partial_\mu A^\mu_-)
+ i g (\partial_\mu A^\nu_- A^\mu_+  -  \partial_\mu A^\nu_+ A^\mu_-)
+ \nonumber \\
& & \makebox[-6cm]{}
 g^2[-A_{\mu,+} A^\mu_- A^\nu_3  +  {1 \over 2} A_{\mu,3}(A^\nu_+ A^\mu_-
 + A^\nu_- A^\mu_+ )] = J^\nu_3
\label{eqm1}
\end{eqnarray}
and
\begin{eqnarray}
\partial^\mu \partial_\mu A^\nu_-  -
\partial^\nu \partial_\mu A^\mu_-
& + & \nonumber \\
& & \makebox[-6cm]{}
 i g A^\mu_3  {\stackrel{\leftrightarrow} {\partial^\nu}} A_{\mu,-}
 + i g (A^\nu_3 \partial_\mu A^\mu_-
        - A^\nu_- \partial_\mu A^\mu_3)
+ 2 i g (\partial_\mu A^\nu_3 A^\mu_-
        - \partial_\mu A^\nu_- A^\mu_3)
+ \nonumber \\
& & \makebox[-6cm]{}
 g^2 [A_{\mu,3} (A^\mu_- A^\nu_3 - A^\mu_3 A^\nu_-)
  + {1 \over 2}  A_{\mu,-} (A^\nu_+ A^\mu_-
- A^\nu_- A^\mu_+)] = J^\nu_-
\;,
\label{eqm2}
\end{eqnarray}
where we use the antisymmetric derivative
$A {\stackrel{\leftrightarrow} {\partial}} B = A(\partial B)
 -(\partial A)B$.
A third equation is the complex conjugate of
Eq.(\ref{eqm2}).

Next we break these equations up into normal
and zero mode components \cite{Kall93}, and look at the equations for each
Lorentz component $\nu=+,-$ and each color component $a=3,+$.
With the above gauge conditions
the $\nu=+$ equations are
\begin{eqnarray}
 (i\partial^+)^2\nm{A^-_3} & = &  \nm{J^+_3}, \\
0 & = & \zm{J^+_3}, \label{3Gauss}  \\
(i\partial^+ + g \zm{A^+_3})^2 \nm{A^-_-} & = & \nm{J^+_-}, \quad {\rm {and}}
\\
g^2(\zm{A^+_3})^2\zm{A^-_-} & =  & \zm{J^+_-}
\;.
\end{eqnarray}
Observe that these equations exhibit no time $\partial_+$ derivatives.
Correspondingly for $\nu = -$:
\begin{eqnarray}
 \partial^+ \partial^- \nm{A^-_3} - {ig \over 2}
             \langle A{^-_-} {\stackrel {\leftrightarrow} {\partial^+}}  A^-_+
\rangle_n
 + g^2 \zm{A^+_3} \langle  A^-_+ A^- _- \rangle_n  & = & \nm{J^-_3},  \\
-(\partial^-)^2 \zm{A^+_3} -{ig \over 2} \langle  A{^- _-}
{\stackrel {\leftrightarrow} {\partial^+}}   A^-_+ \rangle_o
+ g^2 \zm{A^+_3} \langle A^-_+ A^-_- \rangle_o & = & \zm{J^-_3},   \\
-\partial^+\partial^-\nm{A^-_-}-ig\zm{A^+_3}\partial^-\nm{A^-_-}-
2ig\partial^-\zm{A^+_3}\nm{A^-_-}-ig\langle A^-_3\partial^+ A^-_-\rangle_n + \\
ig
\langle
\partial^+ A^-_3  A^-_-
\rangle_n -g^2
\zm{A^+_3}
\langle A^-_3  A^-_- \rangle_n & = & \nm{J^-_-}, \quad {\rm {and}}   \\
-ig\zm{A^+_3}\partial^-\zm{A^-_-}-
2ig\partial^-\zm{A^+_3}\zm{A^-_-}-ig\langle A^-_3\partial^+ A^-_-\rangle_o + \\
ig \langle \partial^+  A^-_3  A^-_- \rangle_o - g^2 \zm{A^+_3}
\langle  A^-_3  A^-_- \rangle_o & = &  \zm{J^-_-}
\;.
\end{eqnarray}
Note the presence of both constraint and evolution
equations.

The constrained nature of the first set of equations
is not so much a property of the front form, but is rather
the Gauss law exhibiting itself. The equations correspond to the
fact that, in non-covariant gauges, the field $A^-$
is generally a non-dynamical field. In a Hamiltonian approach it plays the role
of a
Lagrange multiplier to the Gauss law.
In the approach we shall take to the quantum theory, we shall implement
these as `strong', namely operator constraints. However, special comment must
be reserved for Eq.(\ref{3Gauss}). It actually does not even occur
since we have gauged away $\zm{A^-_3}$.
If the sources themselves were part of the
dynamical problem then
this equation would have to be reintroduced as a `weak' constraint, namely
applied to physical states of the quantum Hilbert space. In the model
we consider below, the
sources are merely external classical fields, essentially just parameters,
so the specific theory we consider there is only meaningful anyway
if $\zm{J^+_3}$ as a parameter vanishes.

\section{The Zero Mode Source Problem}
\noi
\underline{The Classical Solution.}
We now consider a regime of the theory excited by
sources that are purely time-dependent. The reader is referred
to the final section for more discussion on these sources for this
problem. Vanishing normal mode gluons are then a consistent solution to the
above equations of motion in the normal mode sector.
Only zero mode gluons occur. From the zero mode equations of motion there are
then
only two equations with non-trivial content. The last of the $\nu = +$
equations is
simply solved to give
\be
\zm{A^-_\pm} = { { \zm{J^+_\pm} } \over { g^2 (\zm{A^+_3})^2 } }
\label{gausscons}
\;.
\ee
{}From the $\nu = -$ equations we extract only one relevant equation
\be
-(\partial^-)^2 \zm{A^+_3} + g^2 \zm{A^+_3} \zm{A^-_+} \zm{A^-_-}
=
\zm{J^-_3}
\label{dyneq}
\;.
\ee
We observe that the pure glue theory in 1+1 dimensions involves
only a single genuine degree of freedom, the field $\zm{A^+_3}$.
Substituting our solutions Eq.(\ref{gausscons}) into the dynamical equation
Eq.(\ref{dyneq})
we obtain
\be
-(\partial^-)^2 \zm{A^+_3} + { {\zm{J^+_+} \zm{J^+_-}} \over
{ g^2 (\zm{A^+_3})^3}}  = \zm{J^-_3}
\;.
\ee
{}From this we can see that this reduction of the theory is not equivalent to a
perturbation around the free $(g=0)$ theory.
For convenience we henceforth use the notation
\be
\zm{A^+_3} = v \;, \quad
 x^+ = t \;, \quad
w^2  =  { {\zm{J^+_+} \zm{J^+_-}} \over {g^2} } \quad {\rm{and}} \quad
\zm{J^-_3} = {B\over2}
\;.
\ee
The dynamical equation can then be compactly written as
\be
 -{{\partial}^2 \over \partial t^2} v + {w^2 \over v^3} = {B \over 2}
\label{findyneq}
\;.
\ee
It can be solved by easy reduction to quadrature with solution
\be
\pm it = \int^v {y dy \over \sqrt {By^3 + 2 w^2 G y^2 + w^2}}
\ee
where G is an integration constant.

\noi
\underline{The Solution to the Quantum Problem.}
We pursue a Hamiltonian formulation where, in the front form,
the generator of $x^+$ translations $P^-$ or light-cone energy operator
is taken as the Hamiltonian.
The only conjugate momentum is
\be
p\, \equiv\, \zm{\Pi^-_3} \, = \, \partial^- \!\! \zm{A^+_3}\,  =\, \partial^-
v
\;.
\ee
The Hamiltonian density
$ T^{+ -}\, = \, \partial^- \!\! \zm{A^+_3} \Pi^-_3 - \cal L $
leads to the Hamiltonian
\be
 H =  {1 \over 2} [ {p}^2  +  {w^2 \over v^2} + B v] (2L)
\;.
\label{qmHamil}
\ee
Of course, Hamilton's equations of motion agree with
Eq.(\ref{gausscons}) and Eq.(\ref{dyneq}).
Quantization is achieved by imposing a commutation relation at equal light-cone
time
on the dynamical degree of freedom. Introducing the variable
$ q = 2L v $,
the appropriate commutation relation is
\be
 [q(x^+), p(x^+)] = i
\label{comm}
\;.
\ee
Note that the zero mode $v$ or $q$ satisfies a field theory of one
dimension less than the original field theory.
In 1+1 dimensions the field theoretic problem reduces to quantum mechanics of a
single particle as in Manton's treatment of the Schwinger model in
Refs.\cite{cyl}.
One thus has to solve the Schr\"odinger equation
\be
{1 \over 2} (- {d^2 \over dq^2} + {(2Lw)^2 \over q^2} + {{B q} \over {2L}})\psi
= {\cal E}
\psi,
\label{schrodeq}
\ee
with the eigenvalue ${\cal E} = E/(2L)$ actually being an energy density.

Before proceeding with the solution let us briefly show that exactly the
same structure is obtained beginning in the instant form.
Here we introduce the
periodic boundary conditions on a finite interval of length $2L$ in $x^3$.
The appropriate gauge choice is $\partial_3 A^3_a = 0$ and then
a color rotation can single out the diagonal color component of
$v= \zm{A^3_3}$. Zero modes are of course now defined with respect to
the $x^3$ direction. After the color diagonalization, one can
gauge away $\zm{A^0_3}$ and, by analogy to the above, set all
normal mode sources to zero.
With
\be
F^a_{03} = \partial_0 v \delta_{a3} + g \epsilon_{ab3} A^0_b v,
\ee
one gets $p = - \partial_0 v$ as the only conjugate momentum.
The Hamiltonian is now taken as the generator of translations in $x^0$. Thus
\be
H = {1\over2} [p^2 - {g^2} (\zm{A^0_\alpha})^2 v^2 +
2 \zm{J^0_\alpha} \zm{A^0_\alpha}  + 2 v \zm{J^3_3} ] (2L)
\;,\qquad \alpha = 1,2.
\ee
The Gauss law is
\be
\zm{A^0_\alpha} = {{\zm{J^0_\alpha}}\over{g^2 v^2}},
\ee
which upon substitution into the Hamiltonian yields
\be
H = {1\over2} [ p^2 +  {{(\zm{J^0_\alpha})^2}\over{g^2 v^2}}
+ 2 v \zm{J^3_3} ] (2L)
\;.
\ee
With the same chiral color convention one has $(\zm{J^0_\alpha})^2 =
\zm{J^0_+} \zm{J^0_-}$ and thus obviously the same Hamiltonian as in
Eq.(\ref{qmHamil}).

Let us return to solving the Schr\"odinger equation Eq. (\ref{schrodeq}).
All eigenstates $\psi$ have the quantum numbers of the naive vacuum adopted in
standard front form field theory: all of them
are eigenstates of the light-cone momentum operator $P^+$ with zero eigenvalue.
The true vacuum is now that state with lowest $P^-$ eigenvalue.
In order to get an exactly soluble system we perform one more simplification.
We eliminate the source $2B = \zm{J^-_3}$.
One of the solutions to Eq.(\ref{schrodeq}) is then
$ \psi(q) = \sqrt {q}\, Z_{\nu} (\sqrt {2 {\cal E}} q)$
where, in the notation of \cite{Grad80}, $Z_\nu$ is the Bessel function with
$\nu^2 \equiv (2Lw)^2 + 1/4.$
Note that $wL$ is independent of $L$ if $w$,
which is proportional to the external
source, scales in $L$ like a dynamical source \cite{Dem94}.
The general solution is a
superposition of the regular and irregular Bessel functions, that is
\be
\psi (q) = R\sqrt{q} J_{\nu} (\sqrt{2 {\cal E}} q) +S \sqrt {v} J_{-\nu}(\sqrt
{2 {\cal E}} q)
\;.
\ee
The constants $R$ and $S$ need to be specified by boundary conditions,
square-integrability and continuity of the first derivative.
When $\nu > 1/2$ square integrability leads to $S=0$.
The boundary
condition that is to be imposed comes from the treatment of the Gribov problem.
Since the wave function vanishes at $q=0$ we must
demand that the wavefunctions vanish at the first Gribov horizon $q = \pm 2\pi
/ g$.
The overall constant $R$ is then fixed by normalization.
Note that this requirement does not automatically ensure that the wavefunction
vanishes at all horizons with arbitrary sources present. Therefore the
pieces of the wavefunction for each Gribov region will not be exact
copies of each other.
For the source free case the wavefunctions for the different
regions are indeed exact copies \cite{Het93}.
 The most important feature is the consequence
of the boundary condition at the Gribov horizon. This leads to
the energy density only assuming the discrete values
\be
 {\cal E}_m^{(\nu)} = {g^2 \over {8 {\pi}^2}} (X_m^{(\nu)})^2
, \quad   m = 1,2, \dots,
\label{Evals}
\ee
where $X_m^{(\nu)}$ denotes the m-th zero of the $\nu$-th Bessel function
$J_\nu$.
In general, these zeroes can only be obtained numerically.
Thus
\be
\psi_m (q) = R\sqrt{q} J_{\nu} (\sqrt{2 {\cal E}_m^{(\nu)}} q)
\ee
is the complete solution. The true vacuum is the state of lowest energy namely
with
$m=1$.

\section{Discussion and Perspectives}
Let us first summarize the essential points.
We analyzed pure glue non-Abelian gauge theory in a
compact spatial volume and periodic boundary conditions on the gauge
potentials. Working in the front form Hamiltonian approach,
we demonstrated how one carefully fixes the gauge.
The equations of motion enabled identification of dynamical and constrained
zero mode
variables. We solved
the quantum theory consisting of gluons excited only by pure
time-dependent external sources. This reduction
uncovered a basic regime of non-Abelian gauge theory where the front
and the instant form approaches were seen to be identical. It also reduced
a quantum field theory problem to a quantum mechanical one which could
be solved for the Schr\"odinger representation wavefunction.
With the explicit interaction term for the dynamical zero mode switched off,
we exactly solved the theory in the first Gribov horizon.

The exact solution we obtained is genuinely
non-perturbative in character. It describes vacuum-like states since
for all of these states $P^+=0$. Consequently, they all have zero invariant
mass
$M^2 = P^+ P^-$. The states are labelled
by the eigenvalues of the operator $P^-$. We explain below why the non-zero
sources are useful. But with them non-zero we have obtained a generalization of
the
result of Hetrick \cite{Het93}. The linear dependence on $L$ in the result
for the discrete {\it energy} levels is also consistent with what one would
expect from a loop of color flux running around the cylinder.
In the source-free case Hetrick \cite{Het93} uses a wave function that is
symmetric
about $q = 0$. For our problem this corresponds to
\be
\psi_m(q) = N \cos (\sqrt{2 \epsilon_m} q )
\;.
\ee
where N is fixed by normalization. At the first Gribov horizon $q
=2\pi /g$ and $\psi_m = (-1)^m N $, thus $\sqrt{2 \epsilon_m} 2 \pi /g =m
\pi$ and
\be
\epsilon = {g^2 m^2\over{8}}
\;.
\ee
Note that $ m=1 $ is the lowest energy state and has as expected one node in
the
allowed region $ 0\le g \le 2 \pi /g $. Hetrick \cite{Het93} discusses the
connection
to the results of Rajeev \cite{cyl} and we will not comment here.

For the sources non-zero, the wavefunction
automatically vanishes at the origin since it is made up of the regular part
$J_{+\nu}$. There is thus a discontinuous transition from the source free to
non-free cases. Of course the sources themselves are functions of time, so
that as time evolves they may take the value zero. What this potentially
discontinuous
behaviour under time evolution means remains an open question.
The {\it manifest} equivalence of the front and instant form treatments of this
problem is presumably a consequence of the elimination of all but topological
features and in this respect the topology is identical in the two forms.
In our picture, the two forms will begin to look different
with the introduction of genuine dynamical content.
However, the same physical content should be present.

This calculation offers the lesson that even in a front form
approach, the vacuum might not be just the simple Fock vacuum. Dynamical zero
modes do
imbue the vacuum with a rich structure. However, the advantage of the front
form
is not severely lost. In higher dimensions we expect that
the transverse gluon components are not dynamical
but rather are constrained. If these constraints can be solved, the vacuum
will not be inordinately beyond control. This is in sharp distinction to the
instant form approach.
There is nonetheless one possible scenario in which a simple vacuum could be
restored.
The inclusion of normal mode dynamics via the sources will build additional
states on top of the vacua of the present work. One may be able to consistently
perform subtractions to obtain a true vacuum state with the eigenvalue of
$P^-$ identically zero. When
the naive continuum limit $L \rightarrow \infty$ is taken only the states built
on
the lowest level might remain. This
is still under consideration.

We finish by briefly addressing the program for tackling the
higher dimensional theory, and how our result will actually be
valuable for the problem in 3+1 dimensions.
A crucial observation is that as zero modes are independent of at least
one space coordinate they satisfy a field theory in a fewer
number of space dimensions than the original.
One can thus envisage undertaking a hierarchy of projections from
$3 \rightarrow 2 \rightarrow 1 \rightarrow 0$
space dimensions, at each level extracting
a zero mode theory within the previous
higher dimensional theory. A similar idea lies behind the
recent work of \cite{Dem94}.
In our approach one arrives at a quantum mechanical
problem of similar structure to the one we have solved in the present
work. The difference would be that the dynamical quanta of the
higher dimensional theory  ---
both fermions {\it and} gluons --- will be the sources for the
lower dimensional theory.

Our exact solution with non-vanishing sources provides
for eventual understanding how constrained and other dynamical zero mode quanta
come in at higher dimensions, and how they generate QCD spectroscopy
in the real world of 3+1 dimensions.

\section*{Acknowledgments}
The authors would like to acknowledge J. Bjorken, S. Brodsky, A. Dubin,
St. G\"ullenstern, M. Heyssler, R. Naus, J. Pismak, H. Quinn, D. Robertson
and D. Sch\"utte for
useful comments.
This work was supported in part by grants from the
U.S.Department of Energy and a NATO collaborative grant.
ACK is supported by the DFG under contract
DFG-Gz:Pa 450/1-1 and would like to thank The Ohio State University for its
hospitality. SSP would like to acknowledge
the hospitality of the Stanford Linear Accelerator Center and the
Max-Planck-Institut f\"ur Kernphysik.

\newpage

\section*{Appendix: Notation and Conventions}
The convention for light-cone coordinates we employ is that of \cite{Kog70}
$ x^{\pm} = (x^0 \pm x^3)/\sqrt 2$.
The dot product decomposes as
$A \cdot B = A^+ B^- + A^- B^+$.
Following Dirac \cite{Dir49} , $x^+$ is taken as the time parameter.  The
time derivative is thus $\partial_+ \equiv {\partial}/{\partial}x^+$ and
implied the
metric tensor $g^{\mu \nu}$ leads to $\partial_+ = \partial^-$.
Correspondingly, $\partial_- = \partial / \partial x^- = \partial^+$
is the space derivative.
We consider the theory `compactified' in the space dimension:  the light-cone
space coordinate $x^- \in [-L, +L]$.
Periodic boundary conditions are imposed.  Thus a given field $\phi$ can be
expanded in
Fourier modes where the discrete momenta take values
\be
k^+ = n { \pi \over L},\quad  n = 1, 2, \ldots .
\ee
The missing zero mode $n=0$ is projected out by
\be
\zm{\phi\makebox[0.125cm]{}} \equiv \langle \phi \rangle_0 \equiv {1 \over 2L}
\int_{-L}^{+L}
dx^- \phi(x^-)
\ee
while the sum of the remaining non-zero modes is the normal mode
\be
\nm{\phi\makebox[0.125cm]{}} \equiv
\langle \phi(x^-)\rangle_n \equiv \phi(x^-) - \langle \phi \rangle_0
\;.
\ee
We use the notation of Itzykson and Zuber \cite{Itz85} for writing the
SU(2) gauge theory. The gauge potentials are represented by
\be
A^{\mu} = A{^{\mu} _a} t^a
\;, \; t^a = {i\sigma^a \over 2} \; , \; a = 1, 2, 3
\ee
where $t^a$ are representation matrices satisfying the Lie algebra
\be
  \left[ t^a, t^b \right]   =  -\epsilon^{abc}t^c
\ee
and $\sigma^a$ are the Pauli matrices
\be
 \sigma^1  =  \pmatrix{0&1\cr
1&0},\; \sigma^2 = \pmatrix{0&-i\cr
i&0}, \;  \sigma^3 = \pmatrix{1&0\cr
0&-1}
\;.
\ee
The following identities are useful
\begin{eqnarray}
\sigma^a\sigma^b & = & i \epsilon^{abc}\sigma^c + \delta^{ab}
\\ tr(t^at^b) & = & -1/2 \delta^{ab}
\;.
\end{eqnarray}
In component form, the field strength tensor can be written
\be
F^a_{\mu \nu}  =  \partial_\mu A^a_\nu - \partial_\nu A^a_\mu
 + g \epsilon^{abc} A^b_\mu A^c_\nu
\ee
and
\be
D^\mu_{ab}  =  \partial^\nu \delta_{ab} - g \epsilon_{abc} A^\mu_c
\ee
is the covariant derivative in the adjoint representation.

\end{document}